\documentclass[twocolumn]{aastex631}
\usepackage{amsmath}
\usepackage{amssymb}
\usepackage{graphicx}
\usepackage{enumitem}
\setcitestyle{notesep={; }}
\usepackage{threeparttable}

\shorttitle{JWST View of IR-selected TDEs}
\shortauthors{Masterson et al.}

\begin{document}

\title{JWST's First View of Tidal Disruption Events: Compact, Accretion-Driven Emission Lines \& Strong Silicate Emission in an Infrared-selected Sample}

\author[0000-0003-4127-0739]{Megan Masterson}
\affiliation{MIT Kavli Institute for Astrophysics and Space Research, Massachusetts Institute of Technology, Cambridge, MA 02139, USA}

\author[0000-0002-8989-0542]{Kishalay De}
\affiliation{Department of Astronomy and Columbia Astrophysics Laboratory, Columbia University, 550 W 120th St. MC 5246, New York, NY 10027, USA}
\affiliation{Center for Computational Astrophysics, Flatiron Institute, 162 5th Ave., New York, NY 10010, USA}

\author[0009-0001-9034-6261]{Christos Panagiotou}
\affiliation{MIT Kavli Institute for Astrophysics and Space Research, Massachusetts Institute of Technology, Cambridge, MA 02139, USA}

\author[0000-0003-0172-0854]{Erin Kara}
\affiliation{MIT Kavli Institute for Astrophysics and Space Research, Massachusetts Institute of Technology, Cambridge, MA 02139, USA}

\author[0000-0002-1568-7461]{Wenbin Lu}
\affiliation{Departments of Astronomy and Theoretical Astrophysics Center, University of California at Berkeley, Berkeley, CA 94720, USA}

\author[0000-0003-2895-6218]{Anna-Christina Eilers}
\affiliation{MIT Kavli Institute for Astrophysics and Space Research, Massachusetts Institute of Technology, Cambridge, MA 02139, USA}

\author[0000-0002-5063-0751]{Muryel Guolo}
\affiliation{Department of Physics and Astronomy, Johns Hopkins University, 400 N. Charles St., Baltimore, MD 21218, USA}

\author[0000-0002-4410-5387]{Armin Rest}
\affiliation{Space Telescope Science Institute, Baltimore, MD 21218, USA}
\affiliation{Physics and Astronomy Department, Johns Hopkins University, Baltimore, MD 21218, USA}

\author[0000-0001-5231-2645]{Claudio Ricci}
\affiliation{N\'{u}cleo de Astronom\'{i}a de la Facultad de Ingenier\'{i}a, Universidad Diego Portales,
Av. Ej\'{e}rcito Libertador 441, 
Santiago, Chile}
\affiliation{Kavli Institute for Astronomy and Astrophysics, 
Peking University, 
Beijing 100871, People's Republic of China}

\author[0000-0002-3859-8074]{Sjoert van Velzen}
\affiliation{Leiden Observatory, Leiden University, PO Box 9513, 2300 RA Leiden, The Netherlands}


\correspondingauthor{Megan Masterson}
\email{mmasters@mit.edu}

\begin{abstract}

Mid-infrared (MIR) emission from tidal disruption events (TDEs) is a powerful probe of the circumnuclear environment around dormant supermassive black holes. This emission arises from the reprocessing of intrinsic emission into thermal MIR emission by circumnuclear dust. While the majority of optical- and X-ray-selected TDEs show only weak dust echoes consistent with primarily unobscured sight lines, there have been growing efforts aimed at finding TDEs in obscured environments using MIR selection methods. In this work, we present the first JWST observations of 4 MIR-selected TDEs with JWST Mid-Infrared Instrument (MIRI) Medium-Resolution Spectrometer (MRS). Two of these sources show flares in other wavelength bands (one in optical, one in X-ray), while the other two are MIR-only transients. None of these TDEs showed pre-outburst nuclear activity, but all of the MIRI/MRS observations reveal emission lines associated with highly ionized gas in the nucleus, implying ionization from TDE accretion. Additionally, all four sources show silicate emission features around 10 and 18 \micron\, that are much stronger than the features seen in active galactic nuclei (AGN). We suggest that the emitting dust is optically thin to its own emission and show that the MIR spectrum is consistent with emission from optically thin dust in the nucleus. All four sources show an excess at short wavelengths ($\lambda < 8$~\micron), which could arise from a late-time plateau in the intrinsic flare, akin to what is seen in late-time UV observations of unobscured TDEs, although self-consistent dust modeling is required to fully assess the strength of this late-time plateau.

\end{abstract}

\keywords{High energy astrophysics (739)--Infrared astronomy (786)--Supermassive black holes (1663)--Transient sources (1851)--Infrared spectroscopy (2285)}


\section{Introduction} \label{sec:intro}

In the past 50 years, tidal disruption events (TDEs) have gone from a theoretical concept \citep{Hills1975,Rees1988,Evans1989,Cannizzo1990} to an observable phenomenon \citep[for a recent review, see][]{Gezari2021}, becoming one of the leading methods to study the dominant fraction of dormant supermassive black holes (SMBHs). TDEs occur when a star on a nearly radial orbit gets so close to an SMBH that the tidal forces overcome the star's self-gravity. A fraction of the star's mass falls back onto the SMBH, producing a flare of accretion that can be observed across the electromagnetic spectrum. In the late 1990s, the first TDEs were detected with transient soft, thermal X-ray emission, indicative of the formation of an accretion flow \citep[e.g.][]{Bade1996,Komossa1999b,Komossa1999a,Donley2002,Esquej2007}. A few more TDEs were discovered in the early 2000s with time-domain searches in the ultraviolet (UV) band using GALEX \citep[e.g.][]{Gezari2006,Gezari2008,Gezari2009}. The discovery space is now dominated by time-domain optical surveys, which have increased the discovery rate by a factor of 10 already \citep{vanVelzen2011,Gezari2012,vanVelzen2021b,Hammerstein2023,Yao2023} and are expected to increase the rate again by at least a factor of 10 with next generation surveys like the Legacy Survey for Space and Time (LSST) at the Vera Rubin Observatory \citep{Bricman2020}.

These canonical detection methods tend to find TDEs in relatively gas- and dust-poor environments, as evidenced by the low covering factors measured with TDE dust echoes in the mid-infrared \citep[MIR;][]{vanVelzen2016,Jiang2021a}. However, in gas- and dust-rich environments, the nuclear dust may have a larger covering factor, making it more likely that the dust covers the sight line to the nucleus. Dust around accreting SMBHs reprocesses the higher energy optical/UV/X-ray photons into thermal MIR emission \citep{Lu2016,Jiang2016,vanVelzen2016,vanVelzen2021a}, and thus, the MIR may be one of the few ways to detect TDEs in gas- and dust-rich environments. Initial searches for MIR flares in ultraluminous infrared galaxies (U/LIRGs) revealed a handful of TDE-like events \citep[e.g.][]{Mattila2018,Kool2020}, suggesting an elevated rate of TDEs in obscured environments \citep{Reynolds2022}, although many of these showed some level of activity from an active galactic nucleus (AGN) prior to the MIR flare. Other efforts have involved searching for MIR flares in integrated photometry from the NEOWISE mission \citep{Mainzer2014}, using spectroscopically confirmed galaxies from SDSS \citep{Jiang2021b}. This approach revealed many potential TDE candidates, but was also agnostic to past AGN activity, thereby introducing many AGN contaminants into the sample.

To this end, \cite{Masterson2024} recently reported a clean sample of 12 MIR-selected TDEs within roughly 200 Mpc using nearly a decade of data from the NEOWISE mission. The use of novel difference imaging techniques \citep{Zackay2016,De2020} to identify transient emission on top of significant host galaxy light led to the discovery of numerous new transients, including the closest TDE discovered to date that was missed by optical surveys \citep{Panagiotou2023}. After removing sources with AGN-like WISE colors or narrow line ratios, this sample revealed that MIR-selected TDEs occur at comparable rates to optically and X-ray selected TDEs. While precisely constraining the dust covering factor is not possible with two-band WISE photometry, the lack of optical flares in most of these TDEs suggests a relatively high covering factor ($\gtrsim 10$\%) compared to the canonical 1\% value in optically selected TDEs \citep{Jiang2021a}. 

This sample of MIR-selected TDEs can help solve major puzzles in canonical TDE population studies. For example, theoretical work suggests that TDEs may contribute significantly to the growth of the lowest mass SMBHs ($M \lesssim 10^6 \, M_\odot$; \citealt{Stone2017,Rizzuto2023}), but observationally assessing the role of TDEs in SMBH growth is hampered by the difficulty in constraining how much mass is actually accreted during a TDE. In particular, TDEs are expected to radiate away a sizable fraction of the rest mass energy of the incoming star, i.e. on the order of $E \approx \eta M_\odot c^2 \approx 10^{53}$~erg for a radiative efficiency of $\eta = 0.1$. However, typical observations of TDEs can only account 1\% of this energy \citep[e.g.][]{Piran2015,Lu2018}. This ``missing energy problem" could be explained with radiatively inefficient super-Eddington accretion and associated outflows \citep[e.g.][]{Metzger2016}, significant energy released at late times in a long-lasting accretion flow \citep[e.g.][]{Mummery2020,Mockler2021,Mummery2024b}, significant emission in the unobservable extreme UV \citep[e.g.][]{Lu2018,Mummery2020,Wen2020}, or some combination of these solutions \citep[e.g.][]{Mummery2020, Mummery2021,Mummery2024a}. The IR band has a distinct advantage over optical and X-ray surveys when probing the total energy emitted by TDEs: the dust absorption coefficient is nearly constant across UV and optical wavelengths, meaning that dust is sensitive to nearly all of the radiation emitted directly by the TDE. This is particularly important for testing the UV hypothesis and probing this unobservable band, where TDE emission is expected to peak \citep[$T \approx 10^{4-5}$~K;][]{vanVelzen2021b} and which is likely underestimated with a single-temperature blackbody fit to optical/near-UV/X-ray observations \citep[e.g.][]{Mummery2020,Thomsen2022,Guolo2024,Guolo2025}. Thus, by indirectly probing the total UV radiation, the MIR is sensitive to the total energy budget of TDEs and can better assess their contributions to SMBH growth. 

Accurately constraining TDE energetics with MIR observations requires breaking important degeneracies related to the dust physics that cannot be done with only two-band NEOWISE photometry. JWST is uniquely situated to break these degeneracies, including dust composition, density profile, and temperature. This manuscript presents the first JWST MIRI Medium-Resolution Spectrometer \citep[MIRI/MRS;][]{Argyriou2023} observations of TDEs, with observations of 4 of the 12 sources in the volume-limited sample from \cite{Masterson2024}. These sources all show some additional evidence for a TDE and include the closest TDE discovered to date (WTP14adbjsh), an X-ray transient (WTP17aamzew), a source with a late-time broad He~I line in the near-IR in a morphologically disturbed host (WTP18aajkmk), and an optical transient with transient broad H$\alpha$ emission (WTP18aamwpj). To avoid confusion, we will refer to these sources throughout this work by both their WTP name and a nickname based on the above qualities, as given in Table \ref{tab:obs}. The remainder of the paper is structured as follows. We discuss the observations and data reduction procedures in Section \ref{sec:obs}. The nuclear spectra reveal accretion-driven atomic emission lines and strong silicate emission indicative of optically thin dust, which we discuss in Sections \ref{sec:high_ip_lines} and \ref{sec:continuum}, respectively. Finally, the implications of our findings are discussed in Section \ref{sec:discussion}.


\section{Observations \& Data Reduction} \label{sec:obs}

\begin{table*}[t!]
    \centering
    \caption{Details of JWST MIRI/MRS observations}
    \label{tab:obs}
    \begin{tabular}{ccccccccc}
        \hline
        WTP Name & Nickname & Host Galaxy & $z$ & Obs. Date &  $\Delta t$\tablenotemark{a} & Exp. Time\tablenotemark{b} & Pixel Scale\tablenotemark{c} & SNR\tablenotemark{d} \\
        & & & & & [yrs] & [s] & [pc/pixel] & \\
        \hline
        \hline
        WTP14adbjsh & The closest & NGC 7392 & 0.0106 & 2024-06-24 & 9.3 & 666/666/999 & 43 & 500 \\
        WTP17aamzew & The X-ray transient & NGC 2981 & 0.03465 & 2024-05-06 & 6.8 & 4873/4873/12327 & 135 & 340 \\
        WTP18aajkmk & The galaxy merger & MCG-01-07-028 & 0.0287 & 2024-01-26 & 6.4 & 888/888/2231 & 113 & 250 \\
        WTP18aampwj\tablenotemark{e} & The optical transient & KUG 0143+322 & 0.037503 & 2024-01-26 & 5.5 & 666/666/999 & 146 & 370 \\
        \hline
    \end{tabular}
    \raggedright
    \tablenotetext{a}{$\Delta t$ is the time since the disruption, as quotes in Table 1 of \cite{Masterson2024}.}
    \tablenotetext{b}{The first, second, and third number quoted in this column are for grating A, B, and C, respectively.}
    \tablenotetext{c}{We quote the spatial pixel scale in Channel 1, where the pixel scale is 0.196\arcsec/pixel.}
    \tablenotetext{d}{This is the average SNR across the entire wavelength range, computed by taking the mean ratio of flux to flux uncertainty provided by the data reduction pipeline. This does not account for systematic uncertainty in the instrument calibration. }
    \tablenotetext{e}{We note that this source was also an optical transient with the TNS name AT2018gn.} 
\end{table*}

JWST MIRI/MRS provides spectra from 4.9-27.9 \micron\, using 4 channels (channel 1: 4.9-7.65 \micron, channel 2: 7.51-11.70 \micron, channel 3: 11.55-17.98 \micron, channel 4: 17.70-27.9 \micron) split across two detectors. Each channel is split into three different gratings (short--A, medium--B, and long--C), thereby leading to 12 unique spectral cubes. The field of view (FOV) and pixel size increases across successive channels, ranging from a FOV of 3.2 $\times$ 3.7\arcsec\, and pixel size of 0.196\arcsec\, pixel$^{-1}$ for channel 1, up to a FOV of 6.6 $\times$ 7.7\arcsec\, and pixel size of 0.273\arcsec\, pixel$^{-1}$ for channel 4. Similarly, the MIRI/MRS point spread function (PSF) also increases with wavelength, from $\approx 0.3$\arcsec\, at the shortest wavelengths to $\approx 1$\arcsec\, at the longest wavelengths. The spectral resolution ranges from $R \approx 3500$ at the shortest wavelengths in channel 1, to $R \approx 1500$ at the longest wavelengths in channel 4 (\citealt{Labiano2021}; and see Fig. 5 in \citealt{Jones2023}).

We observed four sources from the volume-limited sample with JWST MIRI/MRS through Cycle 2 (Program ID: 3696, PI: K. De). For each source, we obtained a single integration with a 4-point point source dither pattern. As our sources contain significant contributions from the host galaxy beyond the nuclear point source, we also obtained one offset background observation per source, taken directly after the source observation and with the same exposure time as the source observation. Details of each of the observations, including the total exposure time for each grating set-up, are given in Table \ref{tab:obs}.

We reduced the data using the JWST Science Calibration Pipeline\footnote{\url{https://jwst-docs.stsci.edu/jwst-science-calibration-pipeline-overview}} \citep[version 1.14.0;][]{Bushouse2024} and Calibration Reference Data System (CRDS) version 11.17.25 with context ``jwst\_1241.pmap". We applied the standard 3-stage reduction process. First, the uncalibrated data were passed through the \texttt{Detector1} pipeline, which applies detector-level corrections and fits the ramp. The resulting rate files were then processed in \texttt{Spec2}, which accounts for instrumental effects, including flat fielding, flux calibration, stray light, and fringing. The final extraction of both a 1D spectrum and a 3D data cube are performed in the \texttt{Spec3} pipeline. In this step, we performed master background subtraction, which takes the 1D median, sigma-clipped background spectrum from the entire FOV of the background observations and subtracts it off of the science data. We also used the \texttt{Spec3} pipeline to produce the nuclear spectra, using a conical extraction region whose radius varies as a function of wavelength; for this extraction, we use a radius of 1 FWHM of the PSF. This extraction also applies a standard aperture correction that accounts for light lost outside of the central aperture and subtracts off the emission from a background annular region, for which we use the default parameters of the \texttt{extract\_1d} step of the pipeline. Finally, we account for residual fringing effects that occur in the MIRI bandpass by running the \texttt{residual\_fringing} step of the \texttt{Spec3} pipeline on the extracted 1D nuclear spectrum.

\begin{figure*}
    \centering
    \includegraphics[width=\textwidth]{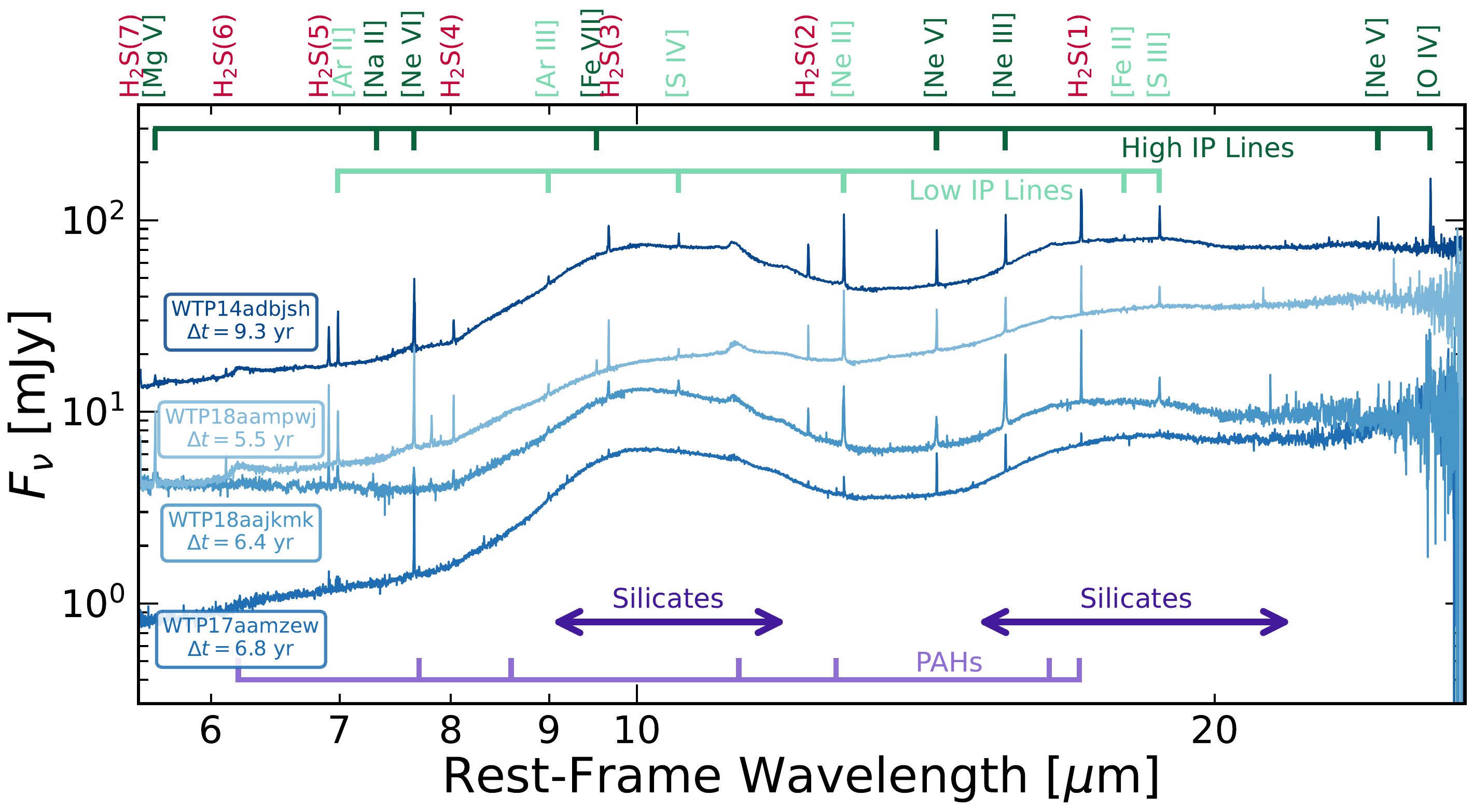}
    \caption{JWST MIRI/MRS nuclear spectra, extracted from within 1 FWHM of the PSF. The four sources in our sample are shown in various shades of blue, varying from darkest for the earliest TDE and lightest for the most recent TDE. The wavelengths of key high and low ionization potential (IP) lines are shown in dark and light green, respectively, where we define the threshold energy between high and low IP at 40 eV. The molecular H$_2$ lines are also labeled on top in red. The location of key silicate features and PAHs are shown in dark and light purple, respectively. All four sources show strong silicate emission at around 10 and 18 \micron, significant emission from the high IP lines associated with black hole accretion, and some, but relatively weak, PAH features. }
    \label{fig:nuclear_spec}
\end{figure*}

The nuclear spectra for each of the four sources in our sample are shown in blue in Figure \ref{fig:nuclear_spec}. All four spectra look remarkably similar, with narrow emission lines from both ionized atomic and molecular gas, broad features emission from silicate dust, and relatively weak emission from polycyclic aromatic hydrocarbons (PAHs). In this work, we focus on the implications of highly ionized atomic lines and strong silicate emission features. The molecular lines, PAH features, and host galaxies will be the subject of a future publication.

\begin{figure*}
    \centering
    \includegraphics[width=\textwidth]{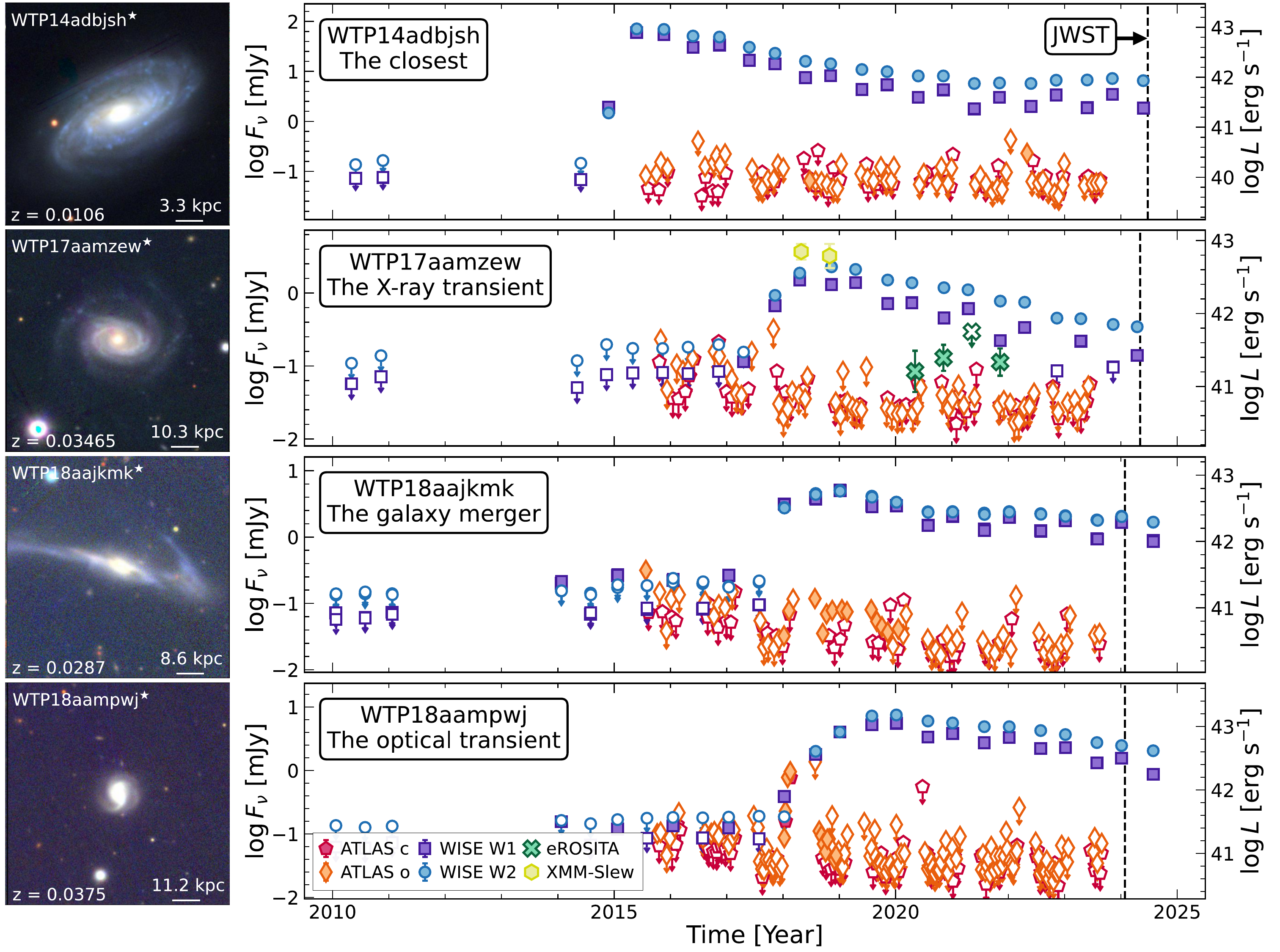}
    \caption{{\it Left:} Host galaxies of the four TDEs in our sample, adapted from Figure 9 in \cite{Masterson2024}. {\it Right:} Multi-wavelength light curves of the four TDEs in our sample, adapted from Figure 1 in \cite{Masterson2024}, including the most recent WISE data up to 2024. The left $y$-axis denotes the flux density in the IR and optical bands. The right $y$-axis denotes the resulting luminosity, corresponding to $\nu L_\nu$ at the central frequency of the $W2$ band for the IR band, the 0.2-2.3 keV luminosity for eROSITA, and the 0.2-2 keV luminosity for XMM-Slew. Red pentagons and orange diamonds denote the ATLAS $c$- and $o$-band data, respectively. The data are binned to 30 days, except for WTP18aampwj (the optical transient), which is binned to 10 day cadence to highlight the optical transient. Green $\times$'s and yellow hexagons show the X-ray data for WTP17aamzew (the X-ray transient), from eROSITA and XMM-Slew, respectively. The vertical black dashed line denotes the time of the JWST observation of each source.}
    \label{fig:nuclear_spec}
\end{figure*}


\section{High-Ionization Emission Lines from Accretion-Driven Flares} \label{sec:high_ip_lines}

We detect many narrow emission lines from ionized gas in the nuclear spectra, with ionization potentials ranging from roughly 10 eV up to nearly 200 eV. All four sources show significant emission from [Ne\,\textsc{vi}] $\lambda$7.65\micron\, with an ionization potential of 126.2 eV, while [Mg\,\textsc{vii}] $\lambda$5.50\micron\, with an ionization potential of 186.5 eV is detected in three of the four sources. The existence of significant emission from these highly ionized species is evidence for black hole accretion, as stellar radiation does not produce enough hard ionizing photons to produce these ionic species with sufficient abundances \citep[see e.g.][]{Abel2008}.

\begin{figure}
    \centering
    \includegraphics[width=\linewidth]{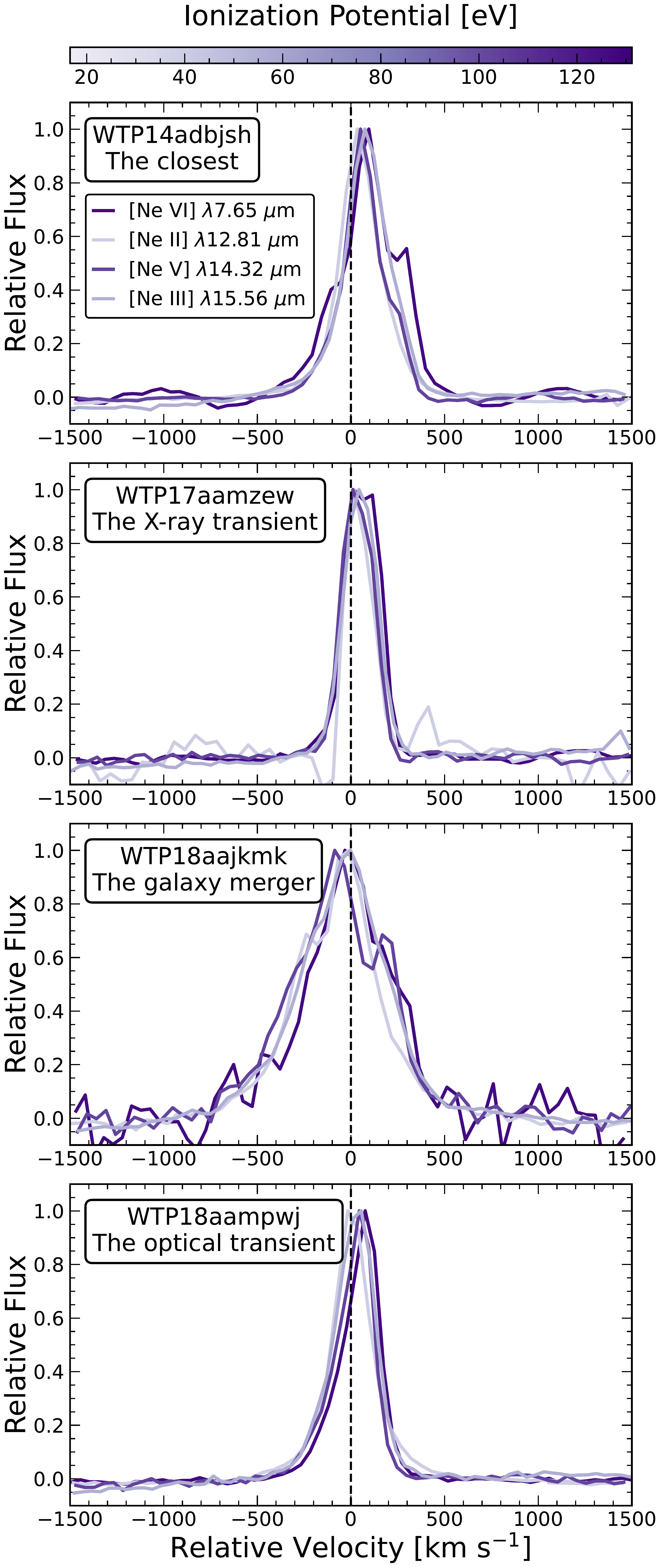}
    \caption{Line profiles for a few of the prominent atomic lines with different IPs. Each panel shows the nuclear spectrum for a different source with lines colored by IP. For each line, we have continuum-subtracted the data, normalized the flux, and shifted to the expected velocity center. We show the [Ne\,\textsc{vi}] $\lambda$7.65\micron\, (IP = 126.2 eV), [Ne\,\textsc{ii}] $\lambda$12.81\micron\, (IP = 26.1 eV), [Ne\,\textsc{v}] $\lambda$14.31\micron\, (IP = 97.1 eV), and [Ne\,\textsc{iii}] $\lambda$15.56\micron\, (IP = 41.0 eV) lines, which span over 100 eV change in IP. In most sources, there is no clear change between profile shapes of the high- and low-ionization lines, with the exception of WTP18aampwj (the optical transient) for which the high IP lines are relatively skewed.}
    \label{fig:line_profs}
\end{figure}

To assess the line profiles, fluxes, and ratios, we fit each emission line of interest within $\pm 1500$ km s$^{-1}$ with a single Gaussian model on top of a linear continuum model with variable slope and intercept. We account for the instrumental width by adding the MIRI MRS line spread function (LSF) from \cite{Argyriou2023} in quadrature with the intrinsic width to produced the observed width \citep[see e.g. Equation (1) in][]{Goold2024}. In Figure \ref{fig:line_profs}, we show the line profiles for a few representative emission lines in velocity space for each source. The lines are colored by the ionization potential (IP) of the ionic species, spanning around 100 eV in IP from primarily star formation driven ionization (e.g. [Ne\,\textsc{ii}] $\lambda$12.81\micron, IP = 26.1 eV) to primarily accretion driven ionization (e.g. [Ne\,\textsc{vi}] $\lambda$7.65\micron, IP = 126.2 eV). The typical widths of these lines are on the order of FWHM $\approx 200-300$~km~s$^{-1}$ in all but WTP18aajkmk (the galaxy merger), which shows larger line widths on the order of FWHM $\approx 700$~km~s$^{-1}$. For the majority of the MIRI bandpass, the resolution ($R \approx 1500-3500$) is sufficient to resolve the lines. Interestingly, in most sources, the ionic lines are slightly redshifted ($v \approx 50-100$~km~s$^{-1}$) with respect to rest frame, while the molecular $H_2$ lines are at the rest frame of the host galaxy. While this could be a result of outflowing gas, there is no significant ($p < 0.01$) trend between the offset velocity nor the velocity dispersion with IP in any of our sources, as has been found and used as evidence for stratified outflows in recent JWST studies of AGN \citep[e.g.][]{Armus2023,HermosaMunoz2024,Zhang2024}. WTP18aampwj (the optical transient) does show evidence for a blue wing in its high IP lines that could be indicative of outflowing gas as was suggested in NGC 7469 \citep{Armus2023}, but we defer a full exploration of the dynamics of the atomic gas to a forthcoming publication.

\begin{figure}
    \centering
    \includegraphics[width=\linewidth]{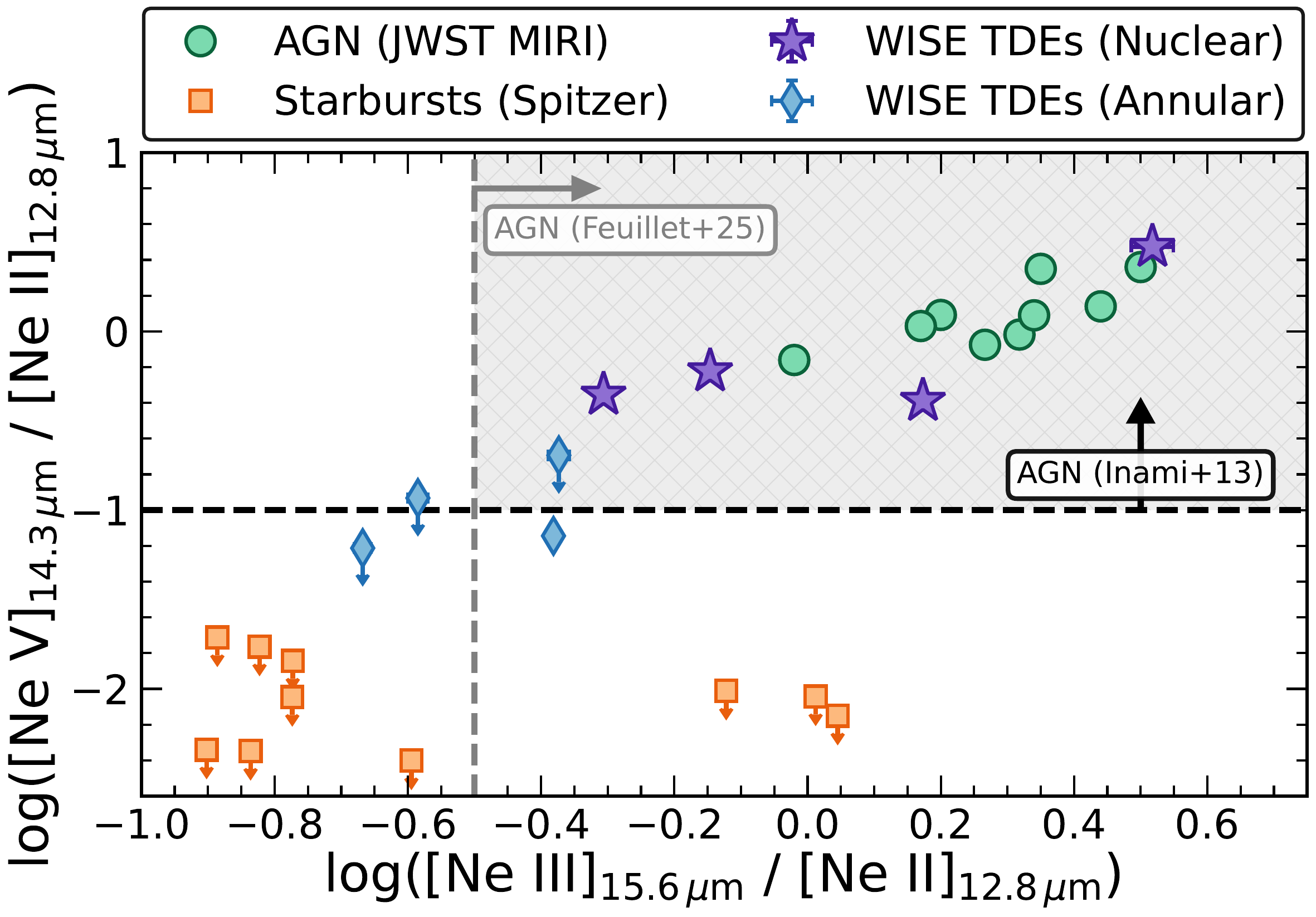}
    \caption{Line ratios for both the nuclear (purple stars) and annular (blue diamonds) MIRI/MRS spectra for each source in our sample as a proxy for the underlying ionizing spectrum. The strength of accretion-driven ionization increases to the upper right part of the plot (i.e. to higher [Ne\,\textsc{v}]/[Ne\,\textsc{ii}] and [Ne\,\textsc{iii}]/[Ne\,\textsc{ii}] ratios). For comparison, the green circles show data from recent JWST MIRI/MRS nuclear spectra of nearby AGN \citep{Pereira-Santaella2022,Armus2023,HermosaMunoz2024,Zhang2024}, and the orange squares show Spitzer IRS high-resolution ($R \sim 600$) data of starburst galaxies, excluding those with a known AGN \citep{Bernard-Salas2009}. The black horizontal and gray vertical lines show different AGN indicators from \cite{Inami2013} and \cite{Feuillet2025}, respectively. The gray hatched region shows the intersection of these two regimes, where the JWST AGN fall. The WISE TDE nuclear spectra show line ratios are comparable to AGN, confirming that their accretion-driven nature. The annular spectra show lower line ratios, suggesting a significant change to the ionization source in the nuclei.}
    \label{fig:line_ratios}
\end{figure}

Line ratios provide critical information about the underlying ionizing spectrum, and, with minimal attenuation from dust, MIR lines are particularly powerful. Many different line ratios across the MIR-FIR spectrum have been used to assess underlying black hole accretion \citep[e.g.][]{Spinoglio1992,Sturm2002,Pereira-Santaella2010,Inami2013,Feltre2023}; for the purposes of this work, we focus on the [Ne\,\textsc{ii}] $\lambda$12.81\micron, [Ne\,\textsc{iii}] $\lambda$15.56\micron, and [Ne\,\textsc{v}] $\lambda$14.32\micron\, lines that are well-suited for MIRI/MRS analysis. Most AGN show high [Ne\,\textsc{v}]~/~[Ne\,\textsc{ii}] and [Ne\,\textsc{iii}]~/~[Ne\,\textsc{ii}] line ratios indicative of a hard ionizing spectrum, while ionization by pure star formation leads to low line ratios, as shown in Figure \ref{fig:line_ratios}. The nuclear spectra of WISE TDEs show comparable line ratios to AGN observed with JWST MIRI/MRS, confirming their accretion-driven nature. It is important to note that these sources are at a similar distance and observed with the same instrument set up, thereby mitigating issues with different aperture sizes and ensuring a fair comparison of the nuclear spectra. Interestingly, the source that shows the highest line ratios is the X-ray transient in our sample (WTP17aamzew), potentially suggesting a uniquely hard ionizing spectrum compared to the other WISE TDEs.

In addition, we extracted a 1D spectrum for each source in an annular region around the nucleus using the 3D data cubes produced during the \texttt{Spec3} pipeline. The inner and outer radius were chosen to be 1.8\arcsec\, ($\approx 0.4-1.3$~kpc, roughly 2 times the FWHM of the PSF at the longest wavelength) and 2.6\arcsec, respectively, to minimize the contributions of the nuclear flux. We applied the additional residual fringing step on these manually extracted spectra. Although these annular spectra are relatively noisy compared to the nuclear spectra, only WTP14adbjsh (the closest) shows weak [Ne\,\textsc{v}] emission and all sources have lower [Ne\,\textsc{iii}] / [Ne\,\textsc{ii}] line ratios (see blue diamonds in Figure \ref{fig:line_ratios}). This implies that there is a significant change to the ionization in the nucleus relative to the rest of the galaxy, where the ionization levels are consistent with being driven by star formation. These high ionization lines also prove that while a small fraction of supernovae can produce long-lived IR flares \citep[e.g.][]{Mo2025}, these nuclear flares with $E_\mathrm{rad} \gtrsim$~few~$\times 10^{51}$~erg are indeed driven by accretion from TDEs, rather than supernovae. For reference, we show the annular spectrum for WTP14adbjsh (the closest) in Appendix \ref{app:annular}.

\section{Continuum Emission \& Strong Silicate Emission Features} \label{sec:continuum}

\subsection{Comparing to AGN \& ULIRGs} \label{subsec:agn_comp}

\begin{figure*}
    \centering
    \includegraphics[width=\textwidth]{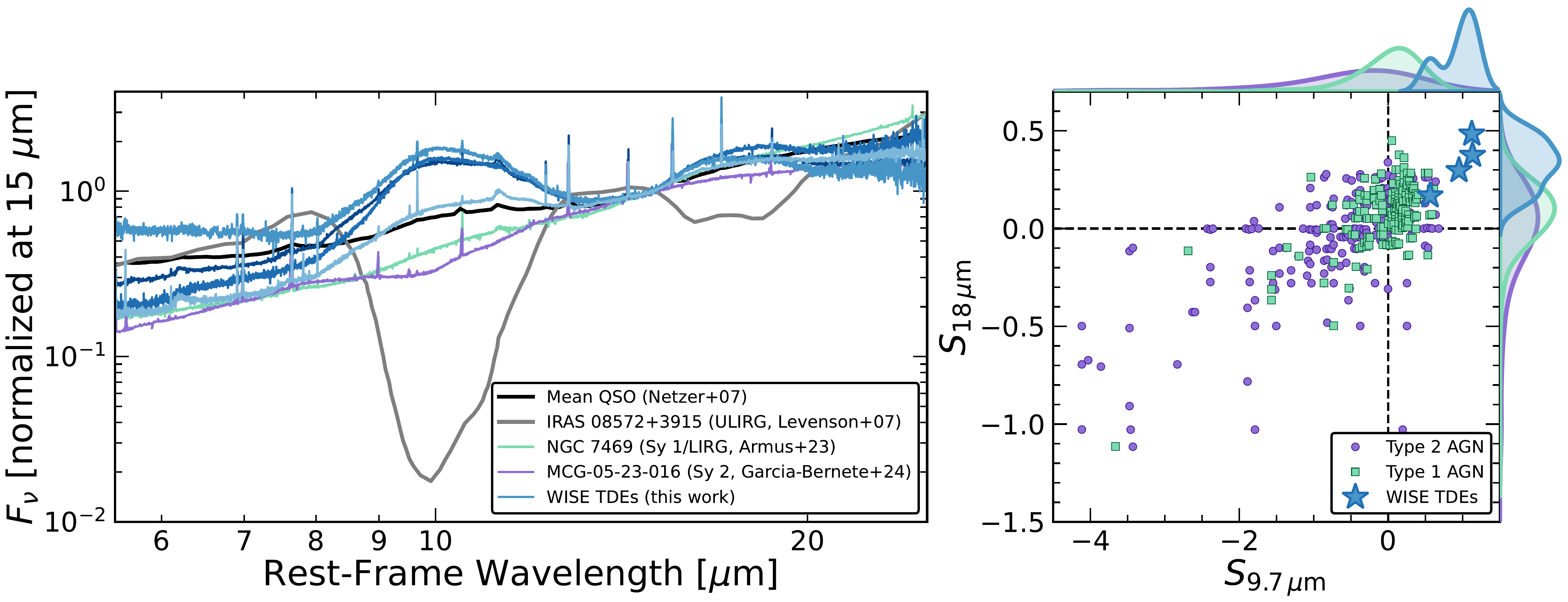}
    \caption{Comparison of TDEs, AGN, and ULIRGs in the MIR band. \textit{Left:} We show the JWST MIRI/MRS nuclear spectra from our TDE sample in blue. The green and purple spectra are examples of Type\,1 and Type\,2 AGN, respectively, at similar distances that have also observed with JWST MIRI/MRS \citep{Armus2023,Garcia-Bernete2024b}. We also show a representative MIR spectrum of a IRAS~08572+3915, a ULIRG observed with the Spitzer Infrared Spectrograph (IRS) with deep silicate absorption in gray \citep{Levenson2007} and the mean QSO Spitzer spectrum from \cite{Netzer2007} in black. \textit{Right:} 9.7 \micron\, vs. 18 \micron\, silicate strengths. A positive (negative) silicate strength corresponds to emission (absorption). The blue stars show our measurements of WISE TDEs with JWST. The green squares and purple circles are Type\,1 and Type\,2 AGN measurements, respectively \citep{Hatziminaoglou2015}. The smoothed histograms on the top and left of this plot show a density of 9.7, 18 \micron\, strengths for each class. The TDEs show significantly stronger silicate emission features than almost all AGN.}
    \label{fig:silicates}
\end{figure*}

In addition to the presence of high-ionization emission lines, one of the most striking features of all four MIR-selected TDE spectra is the presence of strong broad silicate emission at $\sim 9.7, \, 18$ \micron, which arises as a result of stretching and bending of the Si-O and Si-O-Si bonds in silicate dust. Figure \ref{fig:silicates} highlights that the silicate emission features in these TDEs are significantly stronger than most AGN. For a quantitative comparison of these strengths, we employ the canonical definition of $S_\mathrm{sil}$ \citep{Spoon2007}, given by
\begin{equation}
    S_\mathrm{sil} = \ln\frac{f_\mathrm{obs}(\lambda_\mathrm{peak})}{f_\mathrm{cont}(\lambda_\mathrm{peak})},
\end{equation}
where $\lambda_\mathrm{peak}$ is the wavelength that produces the peak of this ratio around both 9.7 and 18 \micron. With this definition, a positive value of $S_\mathrm{sil}$ indicates emission and a negative values indicates absorption. We estimate the continuum by fitting a cubic spline with anchors at 5, 6.25, 7.5, 14, 23, and 25.5 \micron, similar to the procedure detailed in \cite{Spoon2007} but with different points to avoid strong emission lines and stay within the MIRI bandpass. The right panel of Figure \ref{fig:silicates} shows a direct comparison between our TDEs and Type\,1 and Type\,2 AGN from \cite{Hatziminaoglou2015}. Although these comparison data were taken with Spitzer, detailed modeling of the host galaxy emission was removed before measuring the silicate strength, thereby serving as an accurate comparison to the nuclear silicate measurements we make with JWST. 

The strength of these silicate features is directly tied to the geometry and optical depth of the dust, which has been discussed in great detail in the existing literature on both LIRGs and AGN \citep[see e.g.][and references therein]{Sturm2005,Hao2007,Levenson2007,Hatziminaoglou2015}. The deep silicate absorption features in LIRGs require steep temperature gradients in dust with large line-of-sight column densities \citep[e.g.][]{Levenson2007}, while the wide variety of silicate strengths in AGN can naturally be explained by an optically thick, clumpy dust distribution in the torus \citep[e.g.][]{Nenkova2002,Dullemond2005,Nenkova2008a,Nenkova2008b,Schartmann2008,Honig2010}. In the optically thick limit, dust at a single temperature will produce a blackbody spectrum (i.e. $F_\nu \propto B_\nu(T)$), while in the optically thin limit, the emitted spectrum scales as $F_\nu \propto \tau_\nu B_\nu(T)$, where $\tau_\nu$ is the frequency-dependent optical depth. Thus, the existence of strong emission features can naturally be explained by optically thin dust emission.

\subsection{Time-Dependent Optically Thin Dust Modeling} \label{subsec:modeling}

As our observations suggest that the dust around TDEs is optically thin at MIR wavelengths, we proceed with a simple calculation to approximate the dust optical depth within typical galactic centers. Assuming a gas density at the sublimation radius ($r \approx 0.1$~pc) from radio observations of $n_\mathrm{g} \approx 1-100$~cm$^{-3}$ \citep[see Figure 2 in][]{Alexander2020}, dust grains with a typical size of $a \approx 0.1$~\micron\, and bulk density $\rho_\mathrm{bulk} \approx 3$~g~cm$^{-3}$, and a typical gas-to-dust mass ratio of 100 for solar metallicity gas \citep[e.g.][]{Remy-Ruyer2014}, we find a typical dust density of $n_\mathrm{d} \approx 10^{-12} - 10^{-10}$~cm$^{-3}$ for dust in galactic centers. The optical depth is given by
\begin{equation} \label{eq:tau}
    \tau = \pi a^2 \overline{Q}_\nu \int n_\mathrm{d}(r)\, dr,
\end{equation}
where $\overline{Q}_\nu$ is the average absorption efficiency integrated over the source spectrum. In the UV and optical band, $\overline{Q}_\nu \approx 1$, while $\overline{Q}_\nu$ is roughly 2-3 orders of magnitude lower in the IR. For relatively conservative choices of the density profile (e.g. constant density profile with $n_\mathrm{d} = 10^{-9}$~cm$^{-3}$), the dust is indeed optically thin to its own emission for the size scales we are investigating here ($r \sim$~few pc), regardless of the order of magnitude uncertainties on the gas density in galactic centers. 

Motivated by the silicate emission features and this order of magnitude estimate, we modeled the JWST MIRI spectra with optically thin dust, neglecting the geometric effects and scattering, but accounting for the time-dependent effects from TDEs. For this initial investigation, we adopt dust composed of 53\% silicate and 47\% graphite grains \citep[i.e. an MRN mixture, typical for the Milky Way's ISM;][]{Mathis1977} with a single size of $a = 0.1$~\micron. This choice of an MRN dust mixture is motivated by the presence of both silicates and PAHs (i.e. carbon-rich dust) in the MIR spectra. We use the astronomical silicate and graphite optical properties from \cite{Draine1984} and \cite{Laor1993}\footnote{See \url{https://www.astro.princeton.edu/~draine/dust/dust.diel.html}}. To model the underlying flare, we assume that the bolometric luminosity follows the standard optical TDE light curve, and thus, we adopt the post-peak form from Equation (3) in \cite{vanVelzen2021b} with $p = 5/3$ and an average value of $\log_{10} (t_0) = 1.75$ from optical flares \citep{vanVelzen2021b}. We neglect the rising part of the light curve for this initial analysis and simply rescale the peak luminosity of the light curve with a free parameter, $L$. Motivated by recent observations of late-time plateaus in optical and UV data \citep[e.g.][]{Mummery2024b}, we assume that the luminosity remains constant once it reaches a luminosity of $L_\mathrm{plat}$. 

To compute the response of the dust to this flare, we set up a spherically symmetric shell of dust from $r_\mathrm{sub}$ to $r_\mathrm{out}$, approximating the sublimation radius with Equation (1) of \cite{vanVelzen2021a} with the initial flare luminosity and setting $r_\mathrm{out} = 10$~pc. At each point in this shell, we numerically compute the expected temperature of the dust by solving the energy balance equation, given by 
\begin{equation} \label{eq:dust_temp}
    \frac{L(t_r)}{4\pi r^2} \, \overline{Q}_\nu \, e^{-\tau} = \int_0^\infty 4\pi B_\nu\left(T_d\left(r,\theta,t_r\right)\right) Q_\nu \, d\nu,
\end{equation}
where $L$ is the bolometric luminosity emitted by the source at the retarded time given by $t_r = t_\mathrm{obs} - \frac{r}{c}\left(1 - \cos\theta\right)$, $\tau$ is the integrated optical depth given by Equation (\ref{eq:tau}), $Q_\nu$ is the absorption efficiency, and $T_d(r,\theta,t_r)$ is the dust temperature. At the locations where $t_r < 0$, the dust has yet to respond to the flare, and hence there is no transient emission from this dust. This prescription considers only the radiative cooling of the dust, ignoring cooling from grain sublimation, which is only dominant at significantly higher temperatures than probed here ($T \gtrsim 2800$ K; e.g. \citealt{Waxman2000,Lu2016}). Given the unknown source spectrum and the high efficiency of dust absorption at optical and UV wavelengths, we assume $\overline{Q}_\nu = 1$. We adopt a power-law density profile given by $n(r)  = n_\mathrm{sub} (r / r_\mathrm{sub})^{-\gamma}$, where the subscript ``sub" denotes values at the sublimation radius. 

At each point in our shell, we compute the volume emissivity, given by 
\begin{equation}
    J_\nu = \pi a^2 Q_\nu \, n_d(r) \,  B_\nu\left(T_d(r,\theta, t)\right),
\end{equation}
which has units of erg s$^{-1}$ cm$^{-3}$ sr$^{-1}$ Hz$^{-1}$. The total dust emission is then found by integrating the emissivity over the entire shell volume, which yields 
\begin{equation} \label{eq:spectrum}
    F_\nu(t) = \frac{1}{D^2} \int_0^{2\pi} d\phi \int_0^\pi d\theta \int_{r_\mathrm{sub}}^{r_\mathrm{out}} dr \, r^2 \sin\theta \, J_\nu.
\end{equation}
This is akin to the approach presented in \cite{Lu2016}, but with two modifications: (1) we preserve the spectral information, whereas \cite{Lu2016} use Planck-averaged emissivities, and (2) we do not account for the additional attenuation by the outer dust shells (i.e. Equations (18-21) in \citealt{Lu2016}). Intuitively, each point contributes a flux that can be thought of as the solution to the radiative transfer equation for a blackbody source function in the optically thin limit, where $F_\nu(t,\theta,t) \propto B_\nu(T_d(r,\theta,t))\, \tau_\nu$ and depends primarily on the dust temperature.

\begin{figure*}
    \centering
    \includegraphics[width=\textwidth]{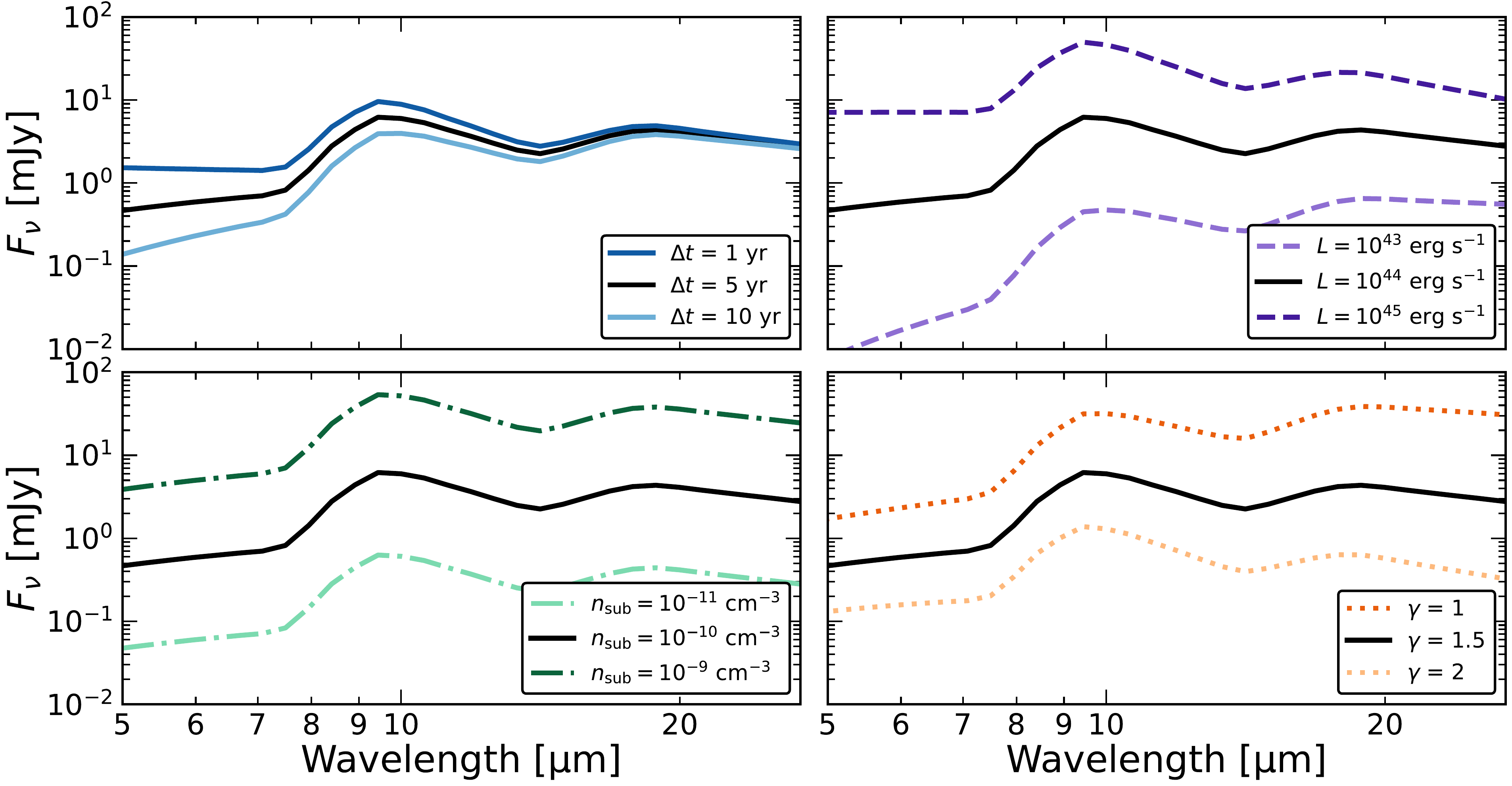}
    \caption{Optically thin dust model for the late-time MIR spectrum from a TDE-like flare. Each panel shows how the model changes for a different free parameter. The free parameters are the time since the flare ($\Delta t$), luminosity of the flare ($L$), density at the sublimation radius ($n_\mathrm{sub}$), and slope of the radial density profile ($\gamma$). The black model is a base model that is the same in each panel. The parameters for the model are: $\Delta t = 5$ yr, $L = 5 \times 10^{44}$~erg~s$^{-1}$, $n_\mathrm{sub} = 10^{-10}$~cm$^{-3}$, and $\gamma = 1$. All models assume a single grain size of $a = 0.1$~\micron, an MRN mixture of silicates and graphite dust, and a plateau luminosity of $L_\mathrm{plat} = 10^{42}$~erg~s$^{-1}$.}
    \label{fig:isodelay_dustmodel}
\end{figure*}

Figure \ref{fig:isodelay_dustmodel} shows how our model depends on each of the free parameters, which are the time since the flare ($\Delta t$), luminosity of the flare ($L$), density at the sublimation radius ($n_\mathrm{sub}$), and slope of the radial density profile ($\gamma$). In all four panels, the black model has $\Delta t = 5$~years, $L = 5 \times 10^{44}$~erg~s$^{-1}$, $n_\mathrm{sub}=10^{-10}$~cm$^{-3}$, $\gamma = 1$, $a = 0.1$~\micron, and $L_\mathrm{plat} = 10^{42}$~erg~s$^{-1}$. Qualitatively, this model matches the late-time TDE spectra from JWST MIRI, with strong silicate emission features and comparable continuum shapes. The spectrum exhibits a clear reddening with time (top left), which is the result of seeing cooler, more distant dust at later times. Both the input luminosity of the flare and $\gamma$ affect the total observed flux and the spectral slope. Increasing the luminosity will lead to higher dust temperatures and therefore a bluer spectrum (top right), while increasing $\gamma$ leads to a more centrally concentrated density profile and a slightly bluer spectrum (bottom right). As long as the dust remains optically thin in the MIR, the density at the sublimation radius (i.e. the normalization) does not appreciably change the spectral slope, just the overall normalization (bottom left). The same is true for the dust grain size, $a$, which we do not show in this figure for brevity. This figure highlights that the parameter space is already quite complex, but the model can provide a qualitatively good explanation of the JWST MIRI spectra.

\begin{figure}
    \centering
    \includegraphics[width=\linewidth]{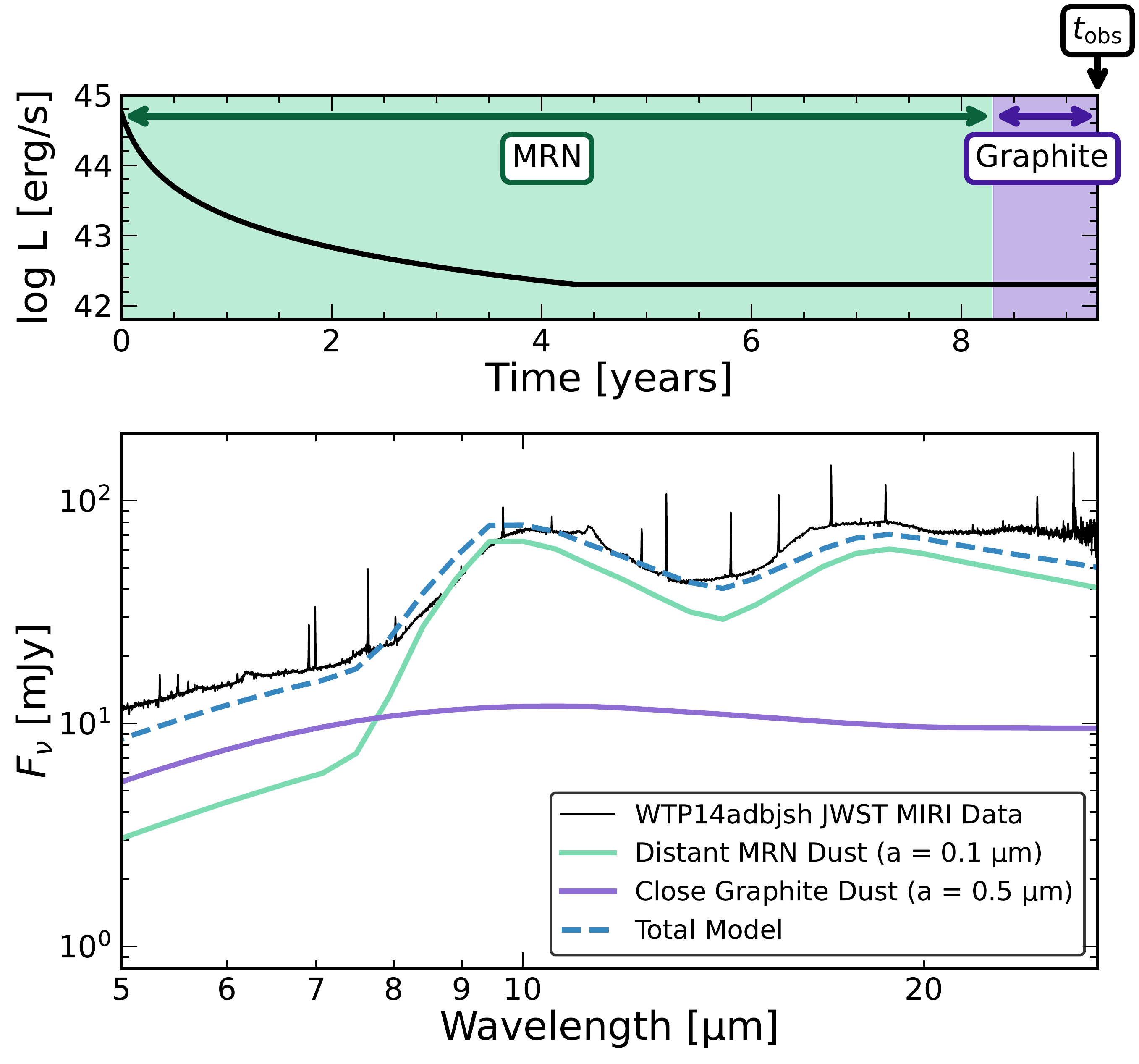}
    \caption{Optically thin dust modeling of WTP14adbjsh (the closest).
    \textit{Top:} Assumed underlying flare heating the dust, with $\Delta t = 9.3$~years, $L = 5.5 \times 10^{44}$~erg~s$^{-1}$, $n_\mathrm{sub} = 1.5 \times 10^{-10}$~cm$^{-3}$, and $\gamma = 0$. The model plateaus when it reaches a luminosity of $L_\mathrm{plat} = 2 \times 10^{42}$~erg~s$^{-1}$. We denote the two phases in which we model the spectrum with MRN and graphite dust in green and purple, respectively. We assume that the graphite dust is responding to the last year of the flare to roughly mimic the dust composition evolution we expect in the innermost regions around the SMBH.
    \textit{Bottom:} Resulting optically thin dust model from the above underlying accretion flare. The MRN dust (green curve) fits the long wavelength spectrum well, but underestimates the short wavelength spectrum. Larger graphite grains near the SMBH (purple curve) can account for this short wavelength excess. We stress that this is not a self-consistent model, nor direct fit to the data, and other parameter combinations could also produce a similarly good fit. We will fully explore this parameter space in a follow-up work with self-consistent dust physics incorporated.} 
    \label{fig:dustmodel_fits}
\end{figure}

Initial comparisons to the JWST MIRI data reveal that our optically thin model provides a good fit in the long wavelength range ($\lambda > 8$~\micron). The bottom panel of Figure \ref{fig:dustmodel_fits} shows an example model for WTP14adbjsh, the closest and oldest TDE in our sample. In this model, the short wavelength flux, which arises primarily from closer regions to the SMBH, is systematically under-predicted with the assumption of MRN dust at a single dust grain size. Modeling the emission from close to the SMBH requires taking into account two important factors: (1) the dust close to the SMBH is likely dominated by graphite because graphite sublimates at a higher temperature than silicates \citep{Barvainis1987}, and (2) smaller dust grains are easier to destroy than larger dust grains \citep{Schartmann2008}, leading to larger dust grains dominating the MIR spectrum in close proximity to the SMBH. While a self-consistent prescription of these two effects is beyond the scope of this work, we show that using larger pure graphite dust grains at late times (i.e. recently heated dust close to the SMBH) provides an adequate description of the short wavelength spectrum. We note that the grain size and plateau luminosity are degenerate, which we will explore in future work, since a fully self-consistent dust model will be necessary to probe this degeneracy accurately. While we only show models for WTP14adbjsh (the closest), these results are similar for all sources in our sample, as Figure \ref{fig:nuclear_spec} suggest that all four sources show a second hot component at short wavelengths. 

\section{Discussion \& Conclusion} \label{sec:discussion}

The exquisite resolving power and angular resolution of JWST MIRI/MRS has allowed us to disentangle the MIR nuclear emission for four nearby ($D_L \lesssim 200$~Mpc) MIR-selected TDEs. These spectra reveal compact, high-ionization emission lines associated with SMBH accretion and strong silicate emission features indicative of optically thin dust emission. Coupled with weak emission from PAHs, the high-IP lines resemble MIR spectra of AGN, rather than star-forming galaxies that show strong PAH features and weak high-IP lines. The similarity to AGN MIR spectra is not particularly surprising, given that both TDEs and AGN are powered by SMBH accretion; the dust around these two systems is heated with similar ionizing radiation that destroys most PAHs and excites atomic gas. However, most AGN show relatively weak silicate features compared to the strong silicate emission features presented in this work. We suspect that this is a direct result of the different dust distributions around active and dormant SMBHs. The canonical clumpy, optically thick torus in AGN is thought to require radiation pressure support from a long-lived accretion flow \citep{Pier1992,Thompson2005,Krolik2007} that does not exist in TDEs. Thus, together with the accretion-driven emission lines, this strengthens the argument that these sources are indeed TDEs.

The discovery of high IP lines in the JWST spectra adds to the growing evidence of TDE-related emission from gas within the central few parsecs of the nucleus.\footnote{While this work was under review, \cite{Mummery2025} used \texttt{CLOUDY} simulations \citep{Ferland1998} to predict that TDEs will, within a matter of months to years, produce bright [Ne\,\textsc{vi}] $\lambda$7.65\micron, [Ne\,\textsc{ii}] $\lambda$12.81\micron, [Ne\,\textsc{v}] $\lambda$14.31\micron, and [Ne\,\textsc{iii}] $\lambda$15.56\micron\, emission lines in the JWST MIRI band with ratios comparable to those of AGN. This result is in agreement with our findings.} For example, observations of nuclear transients have begun to reveal variable [O\,\textsc{iii}] emission, as well as variable emission from iron coronal lines at optical wavelengths \citep[e.g.][]{Wang2012,Hinkle2024b}, suggesting that TDEs in gas- and dust-rich environments can ionize their surroundings, even on short timescales of $\sim$years \citep{Yang2013,Clark2024,Masterson2024,Sanchez-Saez2024}. These lines are tracers of both the surrounding circumnuclear environment and the underlying accretion flare.

Motivated by the strong silicate emission features, we have modeled the JWST MIRI spectra with a simple, optically thin dust model that accounts for the time-dependent spectral evolution. This model can accurately reproduce the strong silicate features and the long wavelength continuum emission. However, the short wavelength emission is dominated by dust near the SMBH heated by the late-time emission from the TDE. Since the initial flare is quite luminous, most of the small grains and silicates at such small radii will have been destroyed. Figure \ref{fig:dustmodel_fits} shows that dust emission from larger ($a = 0.5$~\micron) graphite grains heated by a late-time plateau of $L = 2 \times 10^{42}$~erg~s$^{-1}$ can reproduce the near-IR excess at $\lambda \lesssim 8$~\micron. Interestingly, near-IR excesses are also seen in AGN and have been similarly explained with hot ($T \gtrsim 1200$~K), pure graphite dust that can exist closer to the SMBH than silicate dust \citep[e.g.][]{Mor2009,Mor2012}. As TDEs may emit a sizable fraction of their total energy at late times in a viscously spreading accretion disk \citep{Mummery2020,Mummery2021,Mockler2021,Mummery2024a}, accurate estimation of the late-time plateaus and by association the close dust to the SMBH is crucial to our understanding of TDE energetics and solving the TDE missing energy problem \citep[e.g.][]{Lu2018}. We stress, however, that a more exhaustive search of this complex parameter space is necessary to assess the nature and strength of the late-time plateau in these IR-selected TDEs.

For the purposes of estimating the hot dust emission from the late-time plateau, it is also important to note that we define the innermost dust radius, $r_\mathrm{sub}$, based on the peak luminosity of the flare and assume that all dust destroyed within that region has not yet reformed. Existing AGN literature has demonstrated that dust near the inner edge of the torus may reform on relatively short timescales of $\sim$years \citep[e.g.][]{Koshida2009,Kishimoto2013,Kokubo2020}, likely as a result of new dust condensation in high density ($n \gtrsim 10^{10}$~cm$^{-3}$), neutral gas \citep{Kokubo2020}. However, TDEs do not host the common sites of high density, neutral gas (e.g. the broad line region or atmosphere of the outer accretion disk) assumed in AGN, and the late-time plateau in TDEs can produce significant ionizing radiation on timescales of this observing campaign ($\sim$10 years). It is therefore unclear whether TDEs can actually be a site of rapid dust reformation, thereby justifying our assumption that dust destroyed in the initial flare has not reformed on the timescales of these observations. 

Despite the fact that each of the four sources in our sample shows a unique set of multi-wavelength properties, their JWST MIRI spectra look remarkably similar. This similarity suggests that the dust geometry and optical depth is comparable in these sources and all are unlike standard AGN. Interestingly, the only source with a clear optical flare (WTP18aampwj) is the source with the weakest silicate emission features, suggesting that this source may have the least dust around it, as in the optically thin limit, the strength of these features scales roughly with the optical depth of the dust. This source is also the youngest TDE in our sample ($\Delta t = 5.5$ years), and the silicate feature strengths should also roughly increase with time in the first few years after disruption, as the relative contribution from pure graphite dust decreases like the intrinsic flare luminosity. However, both WTP17aamzew (the X-ray transient) and WTP18aajkmk (the galaxy merger) have stronger silicate features than WTP14adbjsh (the closest), despite both occurring more recently. Both the circumnuclear environment and the time since the TDE impact the resulting MIR spectrum, and thus, continued monitoring of these and other sources at different times with respect to their initial TDE flare will test which of these two effects has the most dominant impact on the resulting MIR spectrum.

Given the complexity associated with astrophysical dust in extreme environments, our modeling has made a number of key assumptions and approximations. One important assumption that we have made is that the dust is optically thin to its own emission (i.e. $\tau_{9.7\, \mu\mathrm{m}} \lesssim 1$). This assumption allows us to neglect scattering in our calculations, as a proper treatment of scattering is difficult to do in a non-spherical emitting surface given geometric effects. Our order of magnitude estimates suggest that optically thin dust is relevant for the timescales (i.e. distances) we are probing with JWST MIRI, but there may be dense, optically thick dust closer to the SMBH. Such dust would impact our calculations by attenuating the dominant UV emission and leading to the more dust being heated by the reprocessed radiation from the inner dust, rather than the intrinsic flare. Future efforts with Monte Carlo radiation transport simulations are necessary to assess the impact of scattering and heating from reprocessed emission on with the necessary time-dependent effects that exist in TDEs. Such simulations are now starting to be viable computationally, but to-date have only been done with gray opacities to estimate the light curve evolution \citep[e.g.][]{Tuna2025}.

These JWST spectra have proven invaluable for assessing the composition and optical depth of the dust in the nuclei of TDE hosts. However, most studies of IR-bright TDEs rely solely on the two-band photometry from NEOWISE, and thus, it is prudent to compare the JWST spectra to the inferred MIR emission from NEOWISE photometry. Fitting the NEOWISE data with either a single temperate blackbody or modified blackbody systematically underpredicts the long wavelength emission in the JWST band. This difference is likely as a result of the multi-temperature nature of the circumnuclear environment; the NEOWISE data is predominantly probing the hotter late-time plateau at short wavelengths, while the JWST spectra are more sensitive to the further, cooler dust. This necessitates further exploration of the IR spectral evolution to facilitate accurate estimate of TDE energetics, which we plan to explore in a future publication by incorporating the NEOWISE evolution into this optically thin dust model.

This modeling makes two key predictions that are relevant for next-generation IR observatories. At very late times ($\gtrsim 100$~years), the illumination of distant, cool dust will lead to far-IR dust echoes that may be detectable with next-generation far-IR observatories like the PRobe far-Infrared Mission for Astrophysics \citep[PRIMA;][]{Moullet2023}. At very early times ($\lesssim 0.5$~years), the spectrum should be dominated by pure graphite dust, as all of the silicates near the SMBH will be destroyed. JWST observations in the first 6 months after a TDE are necessary to test this hypothesis and inform our modeling of the dust close to the SMBH. Although MIR-selected TDE candidate samples are contaminated by SNe at early times, recent work has shown that only a small fraction ($\sim 0.1\%$) of the dominant contaminants, Type\,Ia SNe, are MIR-bright \citep{Mo2025}, thereby making these early-time searches more tractable. Additionally, in the coming years, numerous new observatories will revolutionize time-domain IR studies with greatly improved cadence, including the Nancy Grace Roman Space Telescope \citep{Spergel2015}, the Wide-field INfrared Transient ExploreR \citep[WINTER;][]{Lourie2020}, the PRime-focus Infrared Microlensing Experiment (PRIME\footnote{\url{http://www-ir.ess.sci.osaka-u.ac.jp/prime/index.html}}), and Cryoscope \citep{Kasliwal2025} in the near-IR and NEOSurveyor \citep{Mainzer2023} in the mid-IR. Likewise, SPHEREx \citep{Dore2014}, which launched in March 2025, will produce all-sky near-to-mid-IR spectra every 6 months that can help identify IR-bright TDEs earlier in their evolution. Thus, we expect that catching TDE dust echoes in their early phases of evolution will be possible with the next generation missions. These observations will break degeneracies between dust composition and accretion luminosity, thereby allowing us to answer fundamental questions related to the overall impact of TDEs on black hole growth.\\


\begin{acknowledgments}
We thank the anonymous referee for their helpful comments. We thank Hagai Netzer and Andy Mummery for insightful discussions. This work is based on observations made with the NASA/ESA/CSA JWST. The data were obtained from the Mikulski Archive for Space Telescopes at the Space Telescope Science Institute, which is operated by the Association of Universities for Research in Astronomy, Inc., under NASA contract NAS 5-03127 for JWST. These observations are associated with program \#3696. Support for program \#3696 was provided by NASA through a grant from the Space Telescope Science Institute, which is operated by the Association of Universities for Research in Astronomy, Inc., under NASA contract NAS 5-03127. We acknowledge the support of the National Aeronautics and Space Administration through ADAP grant number 80NSSC24K0663. MG is supported in part by NASA XMM-Newton grant 80NSSC24K1885.

The data presented in this paper were obtained from the Mikulski Archive for Space Telescopes (MAST) at the Space Telescope Science Institute. The specific observations analyzed can be accessed via \dataset[https://doi.org/10.17909/efbj-mf87]{https://doi.org/10.17909/efbj-mf87}. STScI is operated by the Association of Universities for Research in Astronomy, Inc., under NASA contract NAS5–26555. Support to MAST for these data is provided by the NASA Office of Space Science via grant NAG5–7584 and by other grants and contracts.
\end{acknowledgments}

\facilities{JWST MIRI/MRS}

\software{JWST Science Calibration Pipeline \citep{Bushouse2024}, \texttt{astropy} \citep{AstropyCollaboration2013,AstropyCollaboration2018,AstropyCollaboration2022}}


\bibliography{jwst_paper1.bib}


\appendix

\section{Emission Line Fluxes} \label{app:fluxes}

In Table \ref{tab:lines}, we report the fluxes, uncertainties, and upper limits for each of the emission lines of interest in the MIRI bandpass. These are taken from the nuclear spectra shown in Figure \ref{fig:nuclear_spec}.

\begin{table}[ht]
\begin{threeparttable}
\centering
\caption{Line Fluxes and Uncertainties in the MIRI Bandpass}
\begin{tabular}{lccccc}
\hline
Spectral Line & Wavelength [$\mu$m] & WTP14adbjsh & WTP17aamzew & WTP18aajkmk & WTP18aampwj \\
\hline
H$_2$S(8) & 5.05 & 4.7 $\pm$ 0.8 & 0.2 $\pm$ 0.0 & $<$ 15.7 & 2.1 $\pm$ 0.2 \\
{[Fe\,\textsc{ii}]} & 5.06 & $<$ 7.0 & $<$ 0.6 & $<$ 2.3 & $<$ 4.0 \\
{[Fe\,\textsc{ii}]} & 5.34 & 19.0 $\pm$ 0.6 & $<$ 1.0 & $<$ 8.5 & 4.6 $\pm$ 0.6 \\
{[Fe\,\textsc{viii}]} & 5.45 & 6.1 $\pm$ 0.8 & 0.7 $\pm$ 0.1 & $<$ 4.4 & 10.5 $\pm$ 0.4 \\
{[Mg\,\textsc{vii}]} & 5.50 & 6.8 $\pm$ 1.0 & 0.6 $\pm$ 0.1 & $<$ 19.7 & 7.1 $\pm$ 0.6 \\
H$_2$S(7) & 5.51 & 18.2 $\pm$ 1.1 & 0.3 $\pm$ 0.1 & $<$ 2.9 & 9.8 $\pm$ 0.4 \\
{[Mg\,\textsc{v}]} & 5.61 & 7.8 $\pm$ 0.5 & $<$ 0.7 & 8.3 $\pm$ 1.5 & 24.6 $\pm$ 0.4 \\
H$_2$S(6) & 6.11 & 7.0 $\pm$ 0.6 & 0.3 $\pm$ 0.1 & 1.7 $\pm$ 0.4 & 4.5 $\pm$ 0.3 \\
H$_2$S(5) & 6.91 & 47.8 $\pm$ 0.4 & 0.7 $\pm$ 0.1 & 7.0 $\pm$ 0.2 & 20.9 $\pm$ 0.1 \\
{[Ar\,\textsc{ii}]} & 6.99 & 54.6 $\pm$ 0.3 & 0.4 $\pm$ 0.0 & 7.6 $\pm$ 0.4 & 15.8 $\pm$ 0.1 \\
{[Na\,\textsc{ii}]} & 7.32 & 1.6 $\pm$ 0.4 & $<$ 0.5 & $<$ 0.7 & 1.4 $\pm$ 0.3 \\
{[Ne\,\textsc{vi}]} & 7.65 & 123.5 $\pm$ 1.7 & 8.9 $\pm$ 0.1 & 7.7 $\pm$ 0.4 & 43.4 $\pm$ 0.2 \\
H$_2$S(4) & 8.03 & 30.1 $\pm$ 1.3 & $<$ 0.7 & 4.1 $\pm$ 0.4 & 9.9 $\pm$ 0.2 \\
{[Ar\,\textsc{iii}]} & 8.99 & $<$ 15.8 & $<$ 1.3 & 3.7 $\pm$ 0.9 & 4.8 $\pm$ 0.9 \\
{[Fe\,\textsc{vii}]} & 9.53 & $<$ 8.8 & $<$ 1.1 & 2.6 $\pm$ 0.7 & 5.0 $\pm$ 0.7 \\
H$_2$S(3) & 9.66 & 92.3 $\pm$ 2.1 & 0.9 $\pm$ 0.1 & 11.3 $\pm$ 0.8 & 22.4 $\pm$ 0.3 \\
{[S\,\textsc{iv}]} & 10.51 & 24.7 $\pm$ 0.8 & 0.5 $\pm$ 0.1 & 9.0 $\pm$ 0.5 & 5.4 $\pm$ 0.4 \\
H$_2$S(2) & 12.28 & 74.8 $\pm$ 1.1 & 0.4 $\pm$ 0.1 & 6.0 $\pm$ 0.1 & 12.0 $\pm$ 0.2 \\
{[Ne\,\textsc{ii}]} & 12.81 & 119.7 $\pm$ 1.9 & 1.2 $\pm$ 0.1 & 24.6 $\pm$ 0.4 & 45.7 $\pm$ 0.3 \\
{[Ne\,\textsc{v}]} & 14.32 & 71.6 $\pm$ 0.7 & 3.5 $\pm$ 0.0 & 10.1 $\pm$ 0.2 & 20.2 $\pm$ 0.3 \\
{[Ne\,\textsc{iii}]} & 15.55 & 85.4 $\pm$ 2.2 & 3.9 $\pm$ 0.1 & 36.7 $\pm$ 0.5 & 22.6 $\pm$ 0.5 \\
H$_2$S(1) & 17.04 & 156.2 $\pm$ 0.6 & 1.0 $\pm$ 0.0 & 17.0 $\pm$ 0.1 & 22.2 $\pm$ 0.1 \\
{[Fe\,\textsc{ii}]} & 17.94 & 7.7 $\pm$ 0.6 & $<$ 0.7 & $<$ 1.8 & $<$ 6.9 \\
{[S\,\textsc{iii}]} & 18.71 & 53.7 $\pm$ 1.1 & 0.5 $\pm$ 0.1 & 6.2 $\pm$ 0.4 & 13.1 $\pm$ 0.6 \\
{[Ne\,\textsc{v}]} & 24.32 & 34.8 $\pm$ 1.6 & 2.4 $\pm$ 0.3 & 5.5 $\pm$ 1.1 & 6.0 $\pm$ 1.8 \\
{[O\,\textsc{iv}]} & 25.89 & 83.3 $\pm$ 5.3 & $<$ 4.6 & 24.6 $\pm$ 6.9 & $<$ 22.9 \\
\hline
\end{tabular}
\label{tab:lines}
\begin{tablenotes}
\item All fluxes are quoted in units of $10^{-16}$~erg~s$^{-1}$~cm~$^{-2}$. Where quoted, the uncertainties are $1\sigma$, and any line that is not detected at $3\sigma$ significance is given as a $3\sigma$ upper limit.
\end{tablenotes}
\end{threeparttable}
\end{table}

\section{Example Annular Spectrum} \label{app:annular}

\begin{figure}
    \centering
    \includegraphics[width=0.7\textwidth]{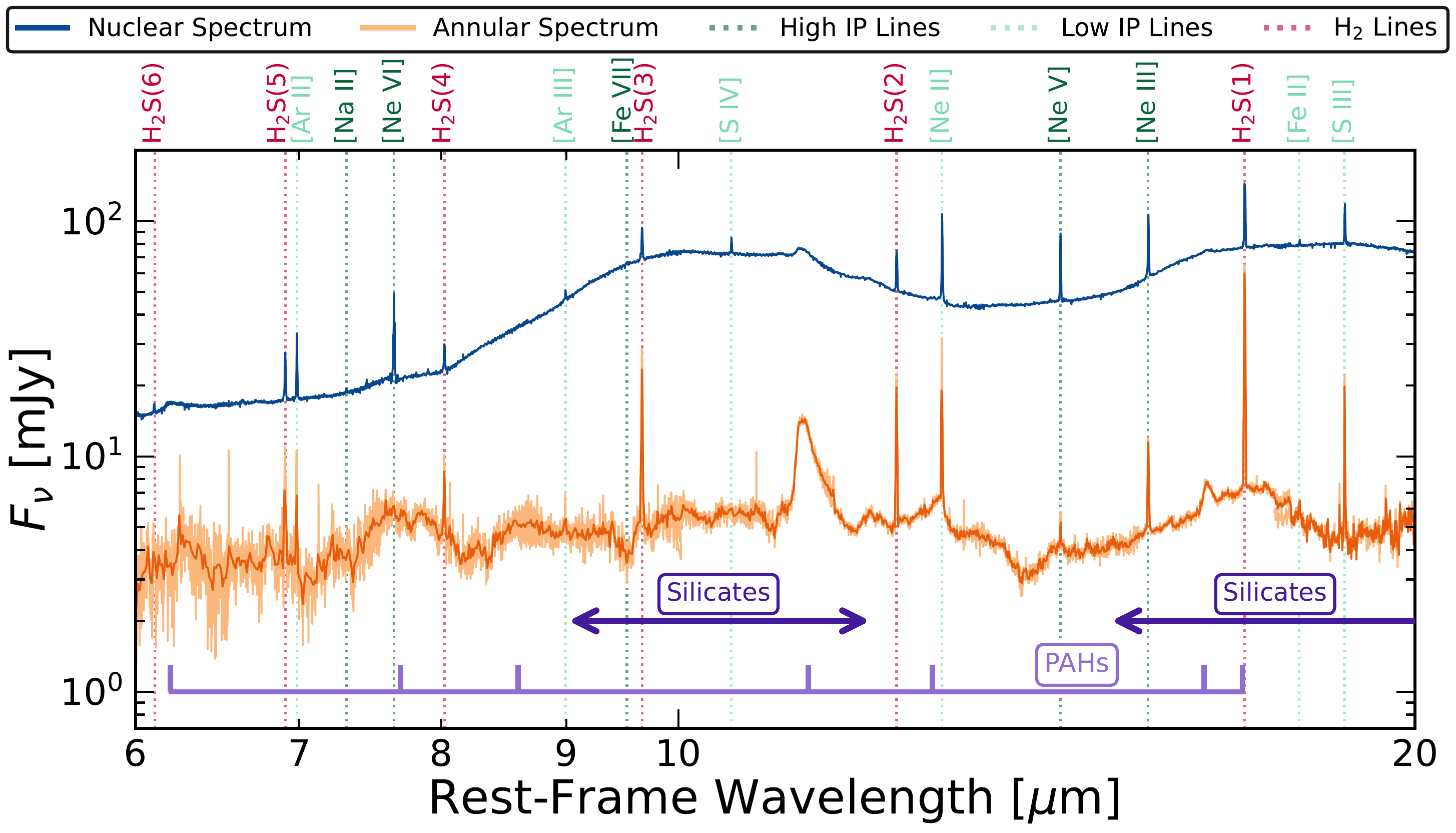}
    \caption{Comparison of JWST MIRI/MRS nuclear and annular spectra for WTP14adbjsh (the closest). The blue spectrum is the nuclear spectrum shown in Figure \ref{fig:nuclear_spec}, with a conical extraction equal to 1 FWHM of the PSF. The annular spectrum is shown in orange, with the darker orange showing the spectrum binned to 0.01 \micron. The wavelengths of key high and low ionization potential (IP) lines are shown in dark and light green, respectively, where we define the threshold energy between high and low IP at 40 eV. The wavelengths of the molecular H$_2$ transitions are shown in red. Likewise, the location of key silicate features and PAHs are shown in dark and light purple, respectively. The PAHs, low IP lines, and molecular lines are much stronger in the annular spectrum compared to the nuclear spectrum.}
    \label{fig:ann}
\end{figure}

In Figure \ref{fig:ann}, we show a comparison between the nuclear and annular spectra of WTP14adbjsh (the closest) in blue and orange, respectively. We show data in the 6-20 \micron\, range, where the annular spectrum is least noisy. The general trends we note below hold for the other three sources, but their annular spectra are noisier as they are dimmer and more distant. The key emission lines associated with accretion (e.g. [Ne\,\textsc{vi}] $\lambda$7.65\micron\,, [Ne\,\textsc{v}] $\lambda$14.32\micron) are much weaker or not detected at all in the annular spectrum, while the molecular H$_2$ lines and low IP lines associated with star formation are much stronger in the annular spectrum. Additionally, the annular spectrum shows significant PAH emission. 

\end{document}